\documentclass[%
reprint,
superscriptaddress,
amsmath,amssymb,
pra,
]{revtex4-2}
\usepackage{url}
\usepackage{braket}
\usepackage{graphicx}
\usepackage{dcolumn}
\usepackage{bm}
\usepackage{color,soul}
\usepackage{comment}
\usepackage{amsmath}
\usepackage{booktabs}
\usepackage{siunitx} 
\usepackage{hyperref}
\usepackage{newfloat}
\DeclareFloatingEnvironment[name={FIG.}]{appFig}

\hypersetup{
colorlinks=true,
linkcolor=blue,
filecolor=magenta,
urlcolor=cyan,
}
\begin{document}

\preprint{APS/123-QED}

\title{Multi-Branch Transport in a Back-gated WS$_2$ Transistor at Deep-Cryogenic Temperature}

\author{Megan Powell}%
\email{Email: megan.powell@strath.ac.uk}
\affiliation{
Department of Physics, SUPA, University of Strathclyde, Glasgow G4 0NG, United Kingdom}
\author{Vilas Patil}%
\affiliation{Tyndall National Institute, University College Cork, Lee Maltings, Dyke Parade, Cork T12 R5CP, Ireland}
\author{Hazel Neill}%
\affiliation{Tyndall National Institute, University College Cork, Lee Maltings, Dyke Parade, Cork T12 R5CP, Ireland}
\author{Stephen O'Sullivan}%
\affiliation{Tyndall National Institute, University College Cork, Lee Maltings, Dyke Parade, Cork T12 R5CP, Ireland}
\author{Paul K. Hurley}%
\affiliation{Tyndall National Institute, University College Cork, Lee Maltings, Dyke Parade, Cork T12 R5CP, Ireland}
\affiliation{School of Chemistry, University College Cork, Cork, Ireland.}
\author{Lida Ansari}%
\affiliation{Tyndall National Institute, University College Cork, Lee Maltings, Dyke Parade, Cork T12 R5CP, Ireland}
\author{Farzan Gity}
\email{Email: farzan.gity@tyndall.ie}
\affiliation{Tyndall National Institute, University College Cork, Lee Maltings, Dyke Parade, Cork T12 R5CP, Ireland}
\author{Alessandro Rossi}
\affiliation{
Department of Physics, SUPA, University of Strathclyde, Glasgow G4 0NG, United Kingdom}
\affiliation{National Physical Laboratory, Hampton Road, Teddington TW11 0LW, United Kingdom
}

\begin{abstract}

Two-dimensional materials are promising candidates for electronic applications beyond the operating limits of conventional semiconductor technologies. Within this class, transition-metal dichalcogenides offer attractive properties for field-effect transistor operation, with tungsten disulphide (WS$_2$) emerging as a particularly promising material for operation at cryogenic temperatures. Here, we investigate the electrical performance of a back-gated multilayer WS$_2$ transistor at deep cryogenic temperature, down to $20~\mathrm{mK}$. The device remains strongly gate-tunable throughout the cryogenic regime, with an effective on/off current ratio exceeding $10^{5}$. Most notably, the low-temperature turn-on characteristics exhibit reproducible shoulder-like features, which we describe using a phenomenological model comprising multiple effective conduction branches operating in parallel. 

\end{abstract}

\maketitle

\section{\label{sec:Intro}Introduction}

Two-dimensional (2D)  semiconductor materials are promising candidates for next-generation electronics beyond conventional silicon. Their atomically thin channels enable strong electrostatic gate control, reduced short-channel effects and leakage, together with compatibility with van der Waals heterostructures~\cite{kim2024future,novoselov2016vdw,sebastian2021benchmarking}. These properties make 2D semiconductors particularly attractive for low-power, high-density transistor technologies requiring efficient switching and stable threshold-voltage control.

Most studies of 2D semiconductor transistors have focused on room-temperature operation, where their potential for scaled logic, flexible electronics, optoelectronics, and low-power integrated circuits has been widely demonstrated~\cite{sebastian2021benchmarking,du2019giant}. However, many emerging technologies require electronic devices to operate under different conditions. For example, the development of cryogenic electronics is becoming increasingly topical for space instrumentation, low-noise sensing, and quantum technologies, where the device under test (DUT) may be required to operate at very low temperature~\cite{vogl2019radiation,liu20192d}. 

At cryogenic temperature, the suppression of thermally activated transport can make the device increasingly sensitive to contact barriers, trapped charge and carrier localisation~\cite{allain2015electrical,schulman2018contact}. These effects can reshape the transistor's turn-on characteristics and alter key circuit-level parameters such as threshold voltage, hysteresis and the DUT's overall current modulation. Understanding how 2D field-effect transistors (FETs) behave under cryogenic operation is therefore important when assessing their suitability for cryogenic and quantum-compatible electronic systems~\cite{olga, das2021transistors}.

Among the most promising 2D semiconductors are transition-metal dichalcogenides (TMDCs), including WS$_2$, whose electronic properties depend strongly on layer thickness and can be readily modified through electrostatic gating~\cite{schaibley2016valleytronics,xu2014spin}. WS$_2$ FETs have demonstrated high on/off current ratios and appreciable carrier mobilities, while retaining compatibility with compact transistor geometries~\cite{schaibley2016valleytronics,xu2014spin,liu2014high}. This combination of electrical tunability and reduced dimensionality makes WS$_2$ particularly interesting for cryogenic transistor architectures and motivates investigation of how its transport characteristics evolve at reduced temperature.

In this work, a back-gated WS$_2$ FET is electrically characterised at cryogenic temperature down to $20~\mathrm{mK}$. The device remains strongly gate-tunable, with a high effective on/off current ratio. Low-temperature operation also reduces significantly gate-voltage hysteresis compared to room-temperature conditions. Importantly, transport measurements reveal reproducible multi-stage turn-on behaviour, which is analysed using a phenomenological multi-branch conduction model. These findings may have important implications for the scalable use of 2D semiconductors in cryogenic technologies, where device operation needs to account for multiple effective threshold voltages rather than assuming a single uniform turn-on point. 

\section{\label{sec:Methods}Methods}

\begin{figure}[h]
\centering
\includegraphics[width=0.5\textwidth]{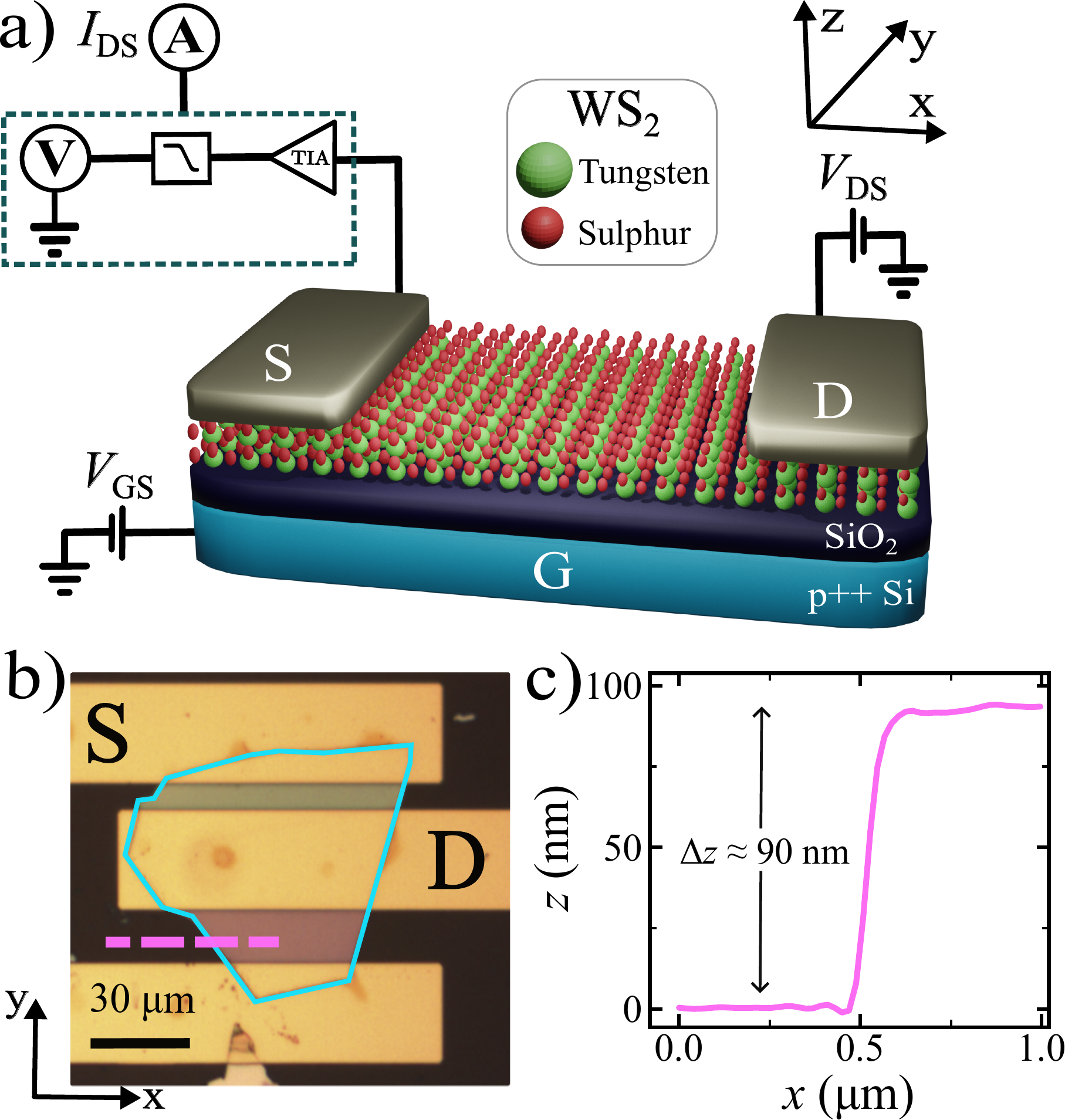}
\caption{ Device structure and physical characterisation of the WS$_2$ flake. (a) Schematic of electrical setup and DUT, consisting of a WS$_2$ flake on a highly doped silicon substrate, acting as a global back G, with a thermally grown oxide layer. S and D contacts are aligned to the flake and defined by metal deposition. (b) Optical microscope image of the WS$_2$ flake, highlighted by a blue contour line, and the source and drain contacts. (c) AFM cross-section indicating a flake thickness of approximately 90 nm along the pink dashed line in (b).}
\label{fig:setup}
\end{figure}

Our DUT, schemtatically shown in Fig~\ref{fig:setup}(a), is a back-gated WS$_2$ FET. The device was fabricated on a highly doped p$^{++}$ Si substrate, which served as the global back gate, with an 85-nm-thick thermally grown SiO$_2$ gate dielectric. WS$_2$ flakes were mechanically exfoliated using the Scotch-tape method~\cite{gao2018mechanical} and transferred onto the Si/SiO$_2$ substrate. Source (S) and drain (D) contacts were then defined lithographically upon the flake, using Ni/Au metallisation with thicknesses of 10/80~nm, respectively. An optical microscope image of the fabricated device is shown in Fig~\ref{fig:setup}(b), where the source and drain contacts are aligned to the exfoliated WS$_2$ flake. The flake thickness was extracted from an atomic force microscopy (AFM) height profile shown in Fig~\ref{fig:setup}(c), giving a thickness of approximately 90~nm.

The device was packaged by wire bonding the chip to a custom built printed circuit board contained in an oxygen free copper enclosure. Temperature-dependent electrical measurements were performed by mounting the packaged device on the mixing-chamber plate of an Oxford Instruments Proteox dilution refrigerator. The plate temperature was controlled using dedicated heaters and measured over the range $20~\mathrm{mK} \leq T \leq 25~\mathrm{K}$, with measurements acquired sequentially from the lowest to the highest temperature setpoint. The sample was maintained under vacuum throughout.

For the main cryogenic measurements, the gate bias, $V_\mathrm{GS}$, was applied to the p$^{++}$ Si back gate using a Keysight B2910BL source-measure unit (SMU), which also monitored the gate leakage current, $I_\mathrm{GS}$, to assess the oxide integrity. The measured leakage was limited by parasitic current paths associated with the long DC wiring loom. Consequently, only a lower bound on the effective leakage resistance could be established, with $R_\mathrm{leak}>1~\mathrm{G\Omega}$. The drain--source bias, $V_\mathrm{DS}$, was applied using an external voltage source, while the drain--source current, $I_\mathrm{DS}$, was measured using a FEMTO DDPCA-300 transimpedance amplifier (TIA). The TIA output voltage was recorded using a Fluke 8842A digital voltmeter and converted to $I_\mathrm{DS}$ in post-processing using the amplifier gain. The adjustable TIA settings allowed the measurement sensitivity to be optimised for different current regimes, with high-gain modes used to resolve small off-state currents and lower-gain settings used where necessary to avoid instrument saturation at larger drain currents. Information about the measurement cycle duration and voltage resolution is reported in Appendix \ref{app:extended_datasets} Table~\ref{tab:loop_duration}.

\section{\label{sec:Results}Results and discussion}

\subsection{Cryogenic Characterisation}
\label{sec:temp_dep}

\begin{figure}[h]
\centering
\includegraphics[width=\linewidth]{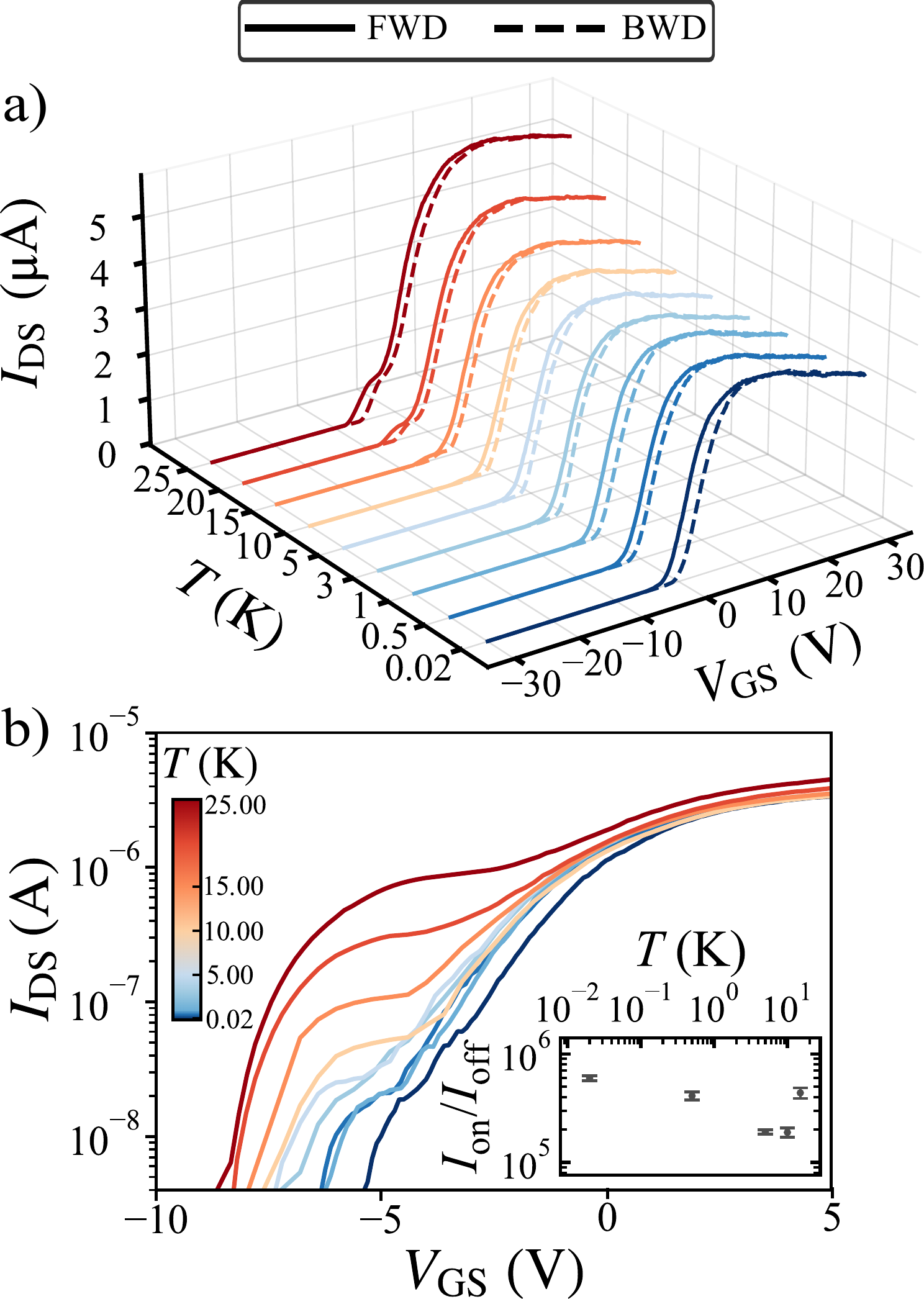}
\caption{Temperature-dependent transfer characteristics of the back-gated WS$_2$ transistor measured over the range $20~\mathrm{mK} \leq T \leq 25~\mathrm{K}$ at $V_{\mathrm{DS}} = 0.5~\mathrm{V}$. (a) Staggered three-dimensional plot of $I_{\mathrm{DS}}$ as a function of gate voltage, $V_{\mathrm{GS}}$, and temperature, $T$. FWD and BWD $V_{\mathrm{GS}}$ sweeps are shown as solid and dashed traces, respectively. (b) FWD-sweep transfer characteristics from panel (a) plotted on a logarithmic current axis, with colour denoting the experimental temperature. The inset shows the extracted effective on/off current ratio as a function of temperature. Error bars in $I_{\mathrm{on}}/I_{\mathrm{off}}$ are obtained by propagating the standard errors of $I_{\mathrm{on}}$ and $I_{\mathrm{off}}$, where $I_{\mathrm{on}}$ is evaluated within a small window around the current maximum and $I_{\mathrm{off}}$ across the OFF-state window.}

\label{fig:transfer_characteristics}
\end{figure}

Figure~\ref{fig:transfer_characteristics}(a) shows the temperature-dependent $I_{\mathrm{DS}}$--$V_{\mathrm{GS}}$ transfer characteristics of the DUT measured between $20~\mathrm{mK}$ and $25~\mathrm{K}$ at a fixed drain--source bias of $V_{\mathrm{DS}} = 0.5~\mathrm{V}$. For each temperature, $V_{\mathrm{GS}}$ was swept from negative to positive bias in the forward (FWD) direction, followed by a sweep from positive to negative bias in the backward (BWD) direction. These sweeps allow the gate-sweep-history dependence of the device response to be assessed. Additional $T = 77~\mathrm{K}$ and room-temperature transfer characteristics are provided in Appendix \ref{app:extended_datasets}.

The device remains strongly gate-tunable across the full measured temperature range, with $I_{\mathrm{DS}}$ increasing as $V_{\mathrm{GS}}$ is made more positive, consistent with n-type field-effect transport in WS$_2$. Importantly, clear gate effect is retained down to $20~\mathrm{mK}$, demonstrating that the device remains electrically active in the millikelvin regime.

Hysteresis is also evident, with a measurable separation between the FWD and BWD gate-voltage sweeps. However, the hysteresis is substantially reduced compared with the room-temperature response (Appendix~\ref{app:extended_datasets}, Fig.~\ref{fig:RT_IV}), indicating improved threshold stability under cryogenic operation. To place this behaviour in context, gate-sweep hysteresis in WS$_2$ and other TMDC transistors is commonly attributed to slow charge trapping and detrapping at the semiconductor/dielectric interface, within the oxide, or at surface-related defect states~\cite{Illarionov2016ChargeTrapping,DiBartolomeo2018MoS2Hysteresis,vanBremen2019WS2Contacts}. Possible contributions to the reduced hysteresis observed here are discussed further in Section~\ref{sec:metrics}.

The transfer curves of Fig~\ref{fig:transfer_characteristics}(a) show non-ideal behaviour in the high-current regime. Rather than forming a fully saturated current plateau at large positive $V_\mathrm{GS}$, the transfer characteristics show a reduction in current at high gate bias - demonstrating negative transconductance. This behaviour is most pronounced under negative drain--source bias, as shown and discussed in the output and $V_\mathrm{DS}$-dependent transfer characteristics provided in Appendix \ref{app:extended_datasets} (see Fig~\ref{fig:output} and~\ref{fig:20mK_Transfer}).

Figure~\ref{fig:transfer_characteristics}(b) shows the FWD-sweep transfer characteristics plotted on a logarithmic current scale as a function of temperature. The transfer curves exhibit a reproducible multi-stage turn-on, with an intermediate-current regime followed by a stronger increase in $I_{\mathrm{DS}}$ at higher gate voltage. As temperature is increased, the current level of the intermediate regime rises, while the gate voltage associated with the stronger turn-on shifts towards lower $V_{\mathrm{GS}}$. This behaviour differs from the more conventional picture of FET turn-on, in which the switch from the OFF state to the ON state occurs via a single sharp transition.

Despite this multi-stage turn-on behaviour, the device retains a clear distinction between its OFF and ON states across the full temperature range. The inset in Fig.~\ref{fig:transfer_characteristics}(b) summarises the effective on/off current ratio, defined as $I_{\mathrm{on}}/I_{\mathrm{off}}$, as a function of temperature. Here, $I_{\mathrm{on}}$ is the maximum $|I_{\mathrm{DS}}|$, while $I_{\mathrm{off}}$ is the mean $|I_{\mathrm{DS}}|$ within a low-conduction gate-voltage window. The ratio remains above $10^{5}$ across the measured temperature range, confirming that the device retains strong gate modulation at cryogenic temperature. This is a significant result, as it demonstrates that the WS$_2$ transistor remains electrically switchable down to millikelvin temperature while maintaining low off-state leakage. The measured on/off ratio is also notably larger than several reported cryogenic TMDC transistor values, including WSe$_2$ devices with on/off ratios of approximately $10^{4}$ at $77~\mathrm{K}$~\cite{kamaei2021gate}. This comparison highlights the strong switching performance of the present WS$_2$ device, despite operation deep in the cryogenic regime.

\subsection{Multi-Branch Transport}
\label{sec:metrics}

A central feature of the measured transfer characteristics is the unusual turn-on behaviour observed in Fig.~\ref{fig:transfer_characteristics}(b). Rather than exhibiting the single smooth transition expected for a uniformly gated channel, the drain current develops a series of distinct shoulder-like features as the gate voltage is increased. To describe this behaviour phenomenologically, we used a multi-branch Lambert-W model~\cite{catapano2023cryogenic}. In this model, each branch represents an effective conduction contribution with its own threshold voltage and branch-strength scaling factor. The fitted drain current was described by

\begin{equation}
\begin{aligned}
I_{\mathrm{DS}} = I_0
&+ \mu |V_{\mathrm{DS}}| V_{\mathrm{T}}
\sum_{m=1}^{n_{\textup{branch}}}
K_{\textup{m}} 
W_0\mathrm{e}^{\left(
\frac{V_{\mathrm{GS}}-V_{\mathrm{th,m}}}{V_{\mathrm{T}}}
\right)} 
\end{aligned}
\label{eq:lambertw_multibranch}
\end{equation}

where $I_{\mathrm{DS}}$ is the fitted drain current and $I_0$ is a fixed offset. $n_{\textup{branch}}$ denotes the total number of effective transport branches included in the fit, while the index $m$ labels the individual branches. The mobility, $\mu$, was fixed during fitting, with $\mu=140~\mathrm{cm^2V^{-1}s^{-1}}$ used for the cryogenic-temperature data, assuming that the mobility is approximately saturated below $T=80~\mathrm{K}$, and $\mu=50~\mathrm{cm^2V^{-1}s^{-1}}$ used for the room-temperature data~\cite{ovchinnikov2014electrical}. The prefactor $K_\textup{m}$ represents an apparent branch-strength scaling factor, defined here as $K_\textup{m}=C_{\mathrm{ox}}(W/L)_\textup{m}$, where $C_{\mathrm{ox}}$ is the oxide capacitance per unit area and $(W/L)_\textup{m}$ is the effective width-to-length ratio associated with branch $m$. Here, $(W/L)_\textup{m}$ is treated as a fitting parameter and is not interpreted as the actual geometry of a distinct physical conduction path. The parameter $V_{\mathrm{th,m}}$ is the threshold voltage of branch $m$. The thermal voltage is given by $V_{\mathrm{T}}=k_{\mathrm{B}}T/q$, where $k_{\mathrm{B}}$ is the Boltzmann constant, $T$ is the measurement temperature, and $q$ is the elementary charge. The notation $W_0$ denotes the principal branch of the Lambert-W function. The transfer characteristics were fitted using an optimisation protocol to assign a total number of branches ($1 \leq n_{\textup{branch}} \leq 3$) at operating temperature.

\begin{figure*}[t]
\centering
\includegraphics[width=\textwidth]{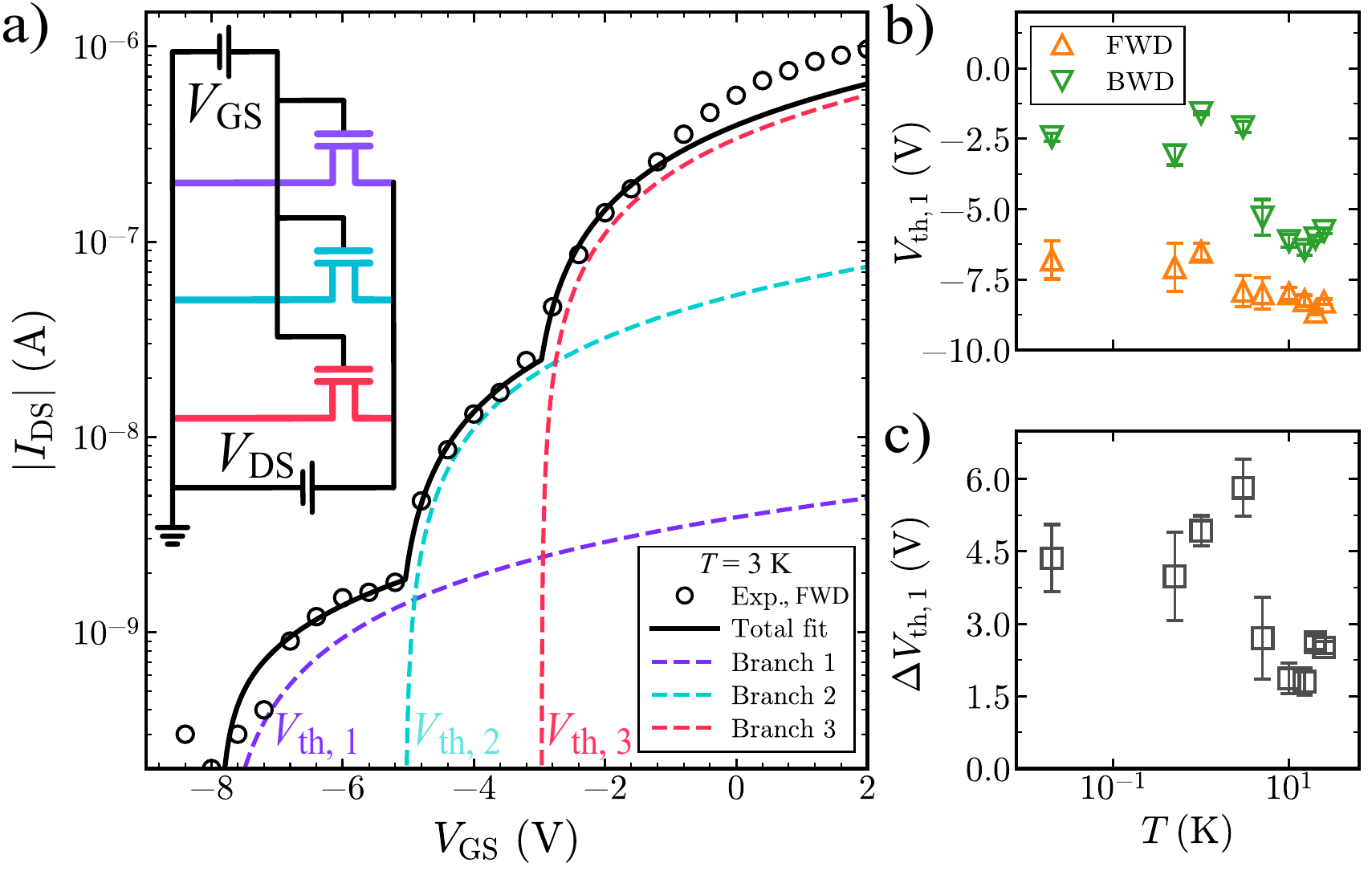}
\caption{Multi-branch Lambert-W fitting analysis of the WS$_2$ transistor transfer characteristics.
(a) Representative fit to the FWD-sweep transfer characteristic at $T = 3~\mathrm{K}$ at $V_\textup{DS}= 0.4 $ V. The measured data are shown as circles, the total fitted current as a solid black line, and the individual branch contributions as dashed coloured lines. The extracted effective branch turn-on voltages, $V_{\mathrm{th},1}$, $V_{\mathrm{th},2}$, and $V_{\mathrm{th},3}$, correspond to the gate voltages at which the separate branch contributions become active. The inset shows a circuit analogy for the multi-branch transport picture, in which the total device current is represented by several effective parallel transistor-like conduction branches. The colour of each schematic branch corresponds to the matching fitted branch in the main plot.
(b) Temperature dependence of the first extracted branch threshold voltage, $V_{\mathrm{th},1}$, for the FWD and BWD gate-voltage sweeps, shown using upward- and downward-pointing triangles, respectively. Error bars represent the $\pm 1\sigma$ uncertainty obtained from the fitting model. All fits are shown for $V_\textup{DS}= 0.4 $ V.
(c) Corresponding first-branch threshold-voltage hysteresis, $\Delta V_{\mathrm{th},1}$, with error bars showing the propagated $\pm 1\sigma$ uncertainty from the extracted FWD and BWD threshold voltages.}
\label{fig:lambertw_fit}
\end{figure*}

An example of the multi-branch fit with three effective branches is shown in Fig.~\ref{fig:lambertw_fit}(a) for the FWD-sweep transfer characteristic measured at $T=3~\mathrm{K}$ with $V_\textup{DS}=0.4~\mathrm{V}$. The corresponding BWD-sweep fit at the same temperature provides a direct comparison between the two sweep directions. The branch number selected by the fitting procedure varies with both temperature and sweep direction, with three branches selected for several of the lowest-temperature FWD sweeps, while fewer branches are generally required as the temperature increases, a trend reiterated in the $77~\mathrm{K}$ and room-temperature measurements. Moreover, the BWD sweeps exhibit a different branch-selection pattern, indicating that the number of resolved contributions depends not only on temperature, but also on the previous gate-bias history of the device. Further details on the corresponding BWD fits, total number of fitted branches, and the room-temperature and $77~\mathrm{K}$ measurements are provided in Appendix~\ref{app:multi-fit} (Figs.~\ref{fig:3K_BWD}, \ref{fig:branch_selection_map} and \ref{fig:MultiTransistor_ExtendedFits} respectively).

The first-branch threshold voltage, $V_{\mathrm{th},1}$, was used as a representative measure of the initial turn-on of the DUT. Fig.~\ref{fig:lambertw_fit}(b) presents the temperature dependence of $V_{\mathrm{th},1}$ for the FWD and BWD gate-voltage sweeps at $V_{\mathrm{DS}}=0.4~\mathrm{V}$. A clear separation between the FWD- and BWD-sweep values is observed across the measured temperature range, with similar trends obtained at $V_{\mathrm{DS}}=0.5$ and $0.6~\mathrm{V}$, as shown in Appendix \ref{app:multi-fit}, Fig.~\ref{fig:extra_vth}. Across the cryogenic range, the extracted thresholds remain within a relatively narrow negative-gate-voltage window and shift gradually toward more negative values as temperature increases. The BWD thresholds are generally less negative than the corresponding FWD thresholds, with a clear FWD--BWD separation retained at every measured temperature, although the magnitude of this separation varies with temperature. The shift of $V_{\mathrm{th},1}$ towards more negative gate voltage with increasing temperature indicates that the effective onset of conduction occurs at a lower applied gate bias. 

Fig.~\ref{fig:lambertw_fit}(c) shows the corresponding threshold-voltage hysteresis, $\Delta V_{\mathrm{th},1}=V_{\mathrm{th},1}^{\mathrm{BWD}}-V_{\mathrm{th},1}^{\mathrm{FWD}}$. A clear hysteresis window is observed at all measured temperatures, confirming that the extracted threshold voltage remains dependent on gate-bias history. Notably, $\Delta V_{\mathrm{th},1}$ remains much smaller throughout the cryogenic range than at room temperature and $77~\mathrm{K}$, as demonstrated in Fig.~\ref{fig:RT_IV}, Appendix~\ref{app:extended_datasets}. The smaller variations observed at the lowest temperatures are likely to be comparable to the experimental uncertainty, reflecting factors such as differences between the mixing-chamber and effective electron temperatures of the DUT, together with the finite gate-voltage step size and fitting uncertainty. Within the cryogenic regime, the hysteresis varies non-monotonically with temperature, indicating that the underlying charge dynamics do not follow a simple thermal dependence.

\subsection{Underlying Physical Mechanisms}

The multi-stage turn-on can be interpreted as conduction being established through several parallel transport contributions that become active over different gate-voltage ranges~\cite{Liu2017VerticalNTC}. This may arise from non-uniform transport within the multilayer WS$_2$ channel, where variations in gate coupling, carrier density, disorder, trap occupation, and interlayer coupling cause different regions or layers to turn on at different gate voltages~\cite{panarella2025impact,Liu2017VerticalNTC}. At cryogenic temperatures, reduced thermal activation can further increase sensitivity to local barriers, disorder, and trapped charge, producing a non-uniform electrostatic landscape and promoting percolative transport~\cite{ghatak2011nature}. The measured current is therefore interpreted as the combined response of several effective conduction pathways with different turn-on voltages. Although their microscopic origin cannot be identified from the transfer characteristics alone, similar percolative transport has been reported in cryogenic and disordered TMDC devices~\cite{paul2016percolative}.

Further support for field-dependent non-uniformity is provided by the complementary DFT calculations in Appendix~\ref{app:comp_methods}. For ideal three- to six-layer WS$_2$ slabs, an applied perpendicular electric field redistributes the band-edge states across the stack, strongly localising the conduction-band minimum towards one outer layer while shifting the valence-band maximum towards the opposite side. While the simplified slabs do not capture the full thickness and screened electrostatics of the experimental device, the calculations provide a microscopic basis for field-induced spatial non-uniformity within multilayer WS$_2$. This behaviour is therefore consistent with spatially non-uniform turn-on under back-gate bias, without requiring a direct assignment of the fitted transport branches to individual layers.

The temperature dependence of the fitted branch structure can also be understood within this picture. The reduction in the number of resolved branches at higher temperature does not necessarily imply the disappearance of the underlying conduction paths; rather, thermal broadening, increased carrier activation, and trap-modified electrostatics may cause neighbouring contributions to merge into a single effective turn-on. The distinct branch-selection behaviour observed between the FWD and BWD sweeps, shown in Appendix~\ref{app:multi-fit}, Fig.~\ref{fig:branch_selection_map}, is also consistent with a role for trapping and detrapping, since the resulting channel electrostatics can depend on the preceding gate-bias history.

A related temperature dependence is observed in the extracted threshold voltage. The shift towards more negative gate voltage with increasing temperature is consistent with increased carrier activation and a corresponding increase in the mobile electron density~\cite{neamen,sze}. At lower temperatures, changes in trap occupation and the freezing of slowly responding charge states may also modify the electrostatic potential experienced by the channel and contribute to the observed threshold response~\cite{jana2023blocking}. For comparison, the room-temperature FWD threshold shown in Appendix~\ref{app:extended_datasets}, Fig.~\ref{fig:RT_IV}, is substantially more negative than the cryogenic values, indicating a greater residual electron population at negative $V_{\mathrm{GS}}$. Although this remains consistent with the n-type field-effect behaviour of WS$_2$, the magnitude of the shift suggests additional contributions from unintentional doping, residual electrostatic charge, or trap-related charge within the WS$_2$/SiO$_2$ structure~\cite{iqbal2016tailoring,neamen}. This comparison therefore supports the conclusion that the cryogenic turn-on behaviour is influenced by both reduced carrier activation and temperature-dependent charge trapping.

The reduced cryogenic hysteresis can be considered within the same framework. This behaviour is consistent with trap-mediated charge dynamics governed by a distribution of activation energies and capture/emission time constants. At cryogenic temperatures, many traps may become frozen or relax only slowly on the measurement timescale, reducing dynamic charge exchange during the gate sweep and thereby suppressing the FWD--BWD threshold separation~\cite{lan2020origin,jana2023blocking}.

\section{\label{sec:Disc}Conclusion}

In summary, a back-gated multilayer WS$_2$ FET was electrically characterised down to $20~\mathrm{mK}$. The device retained strong electrostatic gate control throughout this temperature range, achieving an effective on/off current ratio greater than $10^{5}$ even at millikelvin temperatures. Its turn-on exhibited reproducible multi-stage behaviour, which was described using a multi-branch Lambert-W model. The analysis revealed a temperature-dependent shift of the threshold voltage towards more positive gate bias on cooling, together with a substantial reduction in hysteresis.

Arguably, the threshold-voltage shift, residual hysteresis, contact-limited transport, and multi-branch turn-on behaviour indicate that the measured response is influenced by the contacts, electrostatic disorder, and charge trapping at the WS$_2$/SiO$_2$ interface. Further progress in the reliable cryogenic use of these devices will therefore depend on reducing these extrinsic contributions. Improvements to the dielectric interface and gate stack, together with lower-resistance contacts and cleaner or encapsulated channels, would help stabilise the device response and separate the intrinsic transport properties of WS$_2$ from contact- and trap-related effects \cite{withers2014electron,phan2022enhanced, Xu2016UniversalLowTOhmicContacts}. Moreover, replacing the global back gate with locally defined top-gate structures would provide more spatially selective electrostatic control of the WS$_2$ channel, while also moving the device architecture towards geometries more representative of scalable integrated 2D electronics~\cite{das2021transistors,zeng2024transistor}.

\begin{acknowledgments}

We thank A.~Zotov, E.~Parry and C.~Zhang for useful discussions, as well as J.~Gillan and A.~Robbins for technical support. M.P. acknowledges support from the Mac Robertson Travel Scholarship. A.R. acknowledges support from the UKRI Future Leaders Fellowship Scheme (Grant Agreement UKRI1071). This work was also supported by Research Ireland (formerly Science Foundation Ireland) through the AMBER Research Centre (SFI-12/RC/2278\textunderscore{}P2) and the Frontiers for the Future PI Award (24/FFP-A/13329). The SFI/HEA Irish Centre for High-End Computing (ICHEC) is acknowledged for the provision of computational facilities and support. H.N. and V.P. acknowledge funding through Irish Research Council Award EPSPG/2023/1772 and European Union Marie Sk{\l}odowska-Curie Actions (MSCA) Project 101153933, respectively.

\end{acknowledgments}

\clearpage
\onecolumngrid               

\appendix

\section{Extended datasets}   
\label{app:extended_datasets}

Table~\ref{tab:loop_duration} summarises the $V_\mathrm{GS}$ sweep ranges and corresponding step sizes used for the transfer-characteristic measurements at each temperature outlined in Fig~\ref{fig:transfer_characteristics}. These conditions were applied to each FWD and BWD sweep at a given $V_\mathrm{DS}$. In some measurements, different $V_\mathrm{GS}$ ranges were assigned different step sizes; these are identified as separate regions in the table. Larger step sizes were used within the transistor OFF-state region to reduce the overall measurement time. The final column gives the mean duration of one complete FWD and BWD sweep, including $V_\mathrm{GS}$ stepping, instrument readout and communication, and transfer of the measured data to the PC.

\begin{table*}[h]
\centering
\caption{Gate-sweep conditions and mean duration of one complete FWD-and-BWD hysteresis loop at each temperature. In some cases, two $V_\textup{GS}$ regions with different step sizes were used. Note TIA gain was set to $10^5$ V/A for each temperature, and filtering was 0.7 Hz.}
\label{tab:loop_duration}
\normalsize
\renewcommand{\arraystretch}{1.20}

\begin{tabular*}{\textwidth}
{@{\extracolsep{\fill}} c cc cc c @{}}
    \toprule

    & \multicolumn{2}{c}{$V_{\mathrm{GS}}$ Region 1}
    & \multicolumn{2}{c}{$V_{\mathrm{GS}}$ Region 2}
    & \\

    \cmidrule(lr){2-3}
    \cmidrule(lr){4-5}

    Temperature
    & Range
    & Step
    & Range
    & Step
    & Avg. FWD+BWD duration
    \\

    (K)
    & (V)
    & (V)
    & (V)
    & (V)
    & (min)
    \\

    \midrule

    0.02
    & $-30$ to $+30$
    & 0.100
    & --
    & --
    & 28
    \\

    0.50
    & $-30$ to $+30$
    & 0.120
    & --
    & --
    & 20
    \\

    1.00
    & $-30$ to $+30$
    & 0.120
    & --
    & --
    & 20
    \\

    3.00
    & $-30$ to $-10$
    & 0.667
    & $-10$ to $+30$
    & 0.400
    & 17
    \\

    5.00
    & $-30$ to $+30$
    & 0.200
    & --
    & --
    & 18
    \\

    10.00
    & $-30$ to $-10$
    & 0.667
    & $-10$ to $+30$
    & 0.400
    & 17
    \\

    15.00
    & $-30$ to $-10$
    & 0.667
    & $-10$ to $+30$
    & 0.400
    & 17
    \\

    20.00
    & $-30$ to $+30$
    & 0.200
    & --
    & --
    & 17
    \\

    25.00
    & $-30$ to $-15$
    & 0.150
    & $-15$ to $+30$
    & 0.180
    & 17
    \\

    \bottomrule
\end{tabular*}

\end{table*}

\begin{figure*}[h]
\centering
\includegraphics[width=\textwidth]{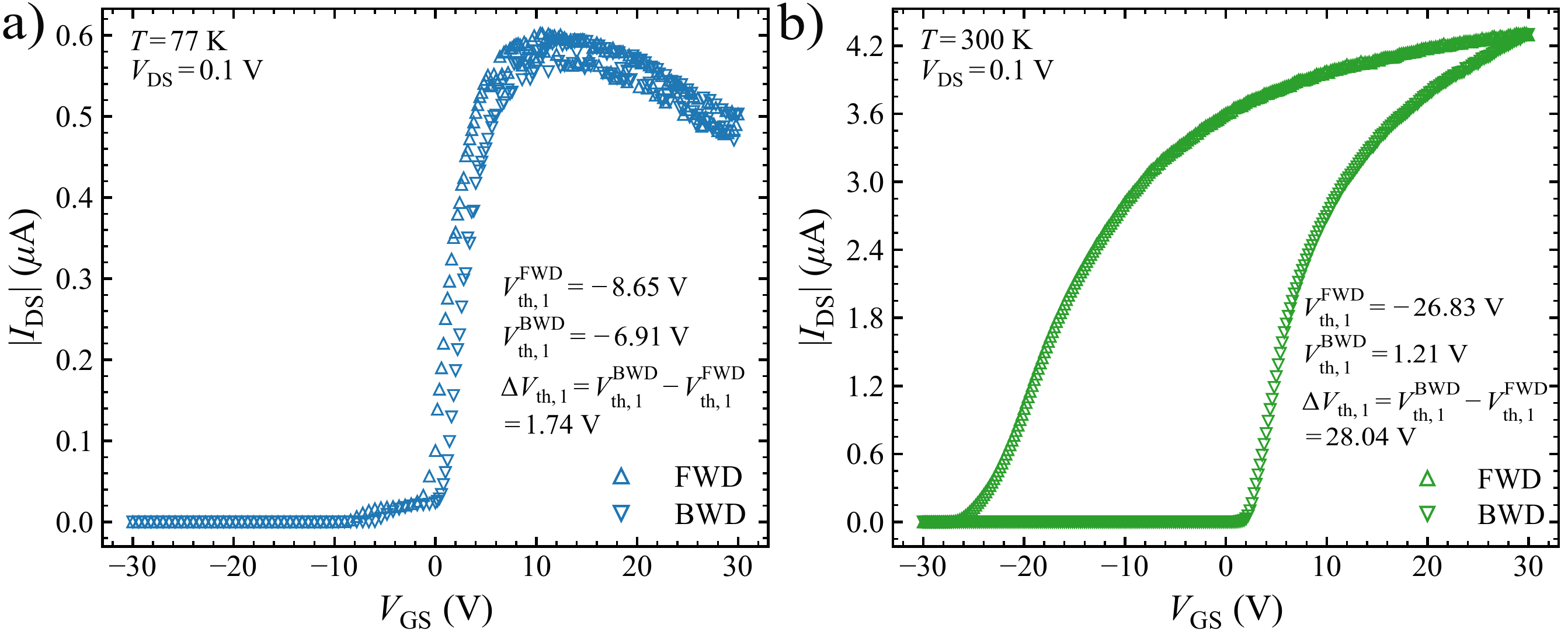}
\caption{Transfer characteristics of the DUT measured at (a) $T = 77~\mathrm{K}$ with $V_{\mathrm{DS}} = 0.1~\mathrm{V}$ and (b) $T = 300~\mathrm{K}$ with $V_{\mathrm{DS}} = 0.1~\mathrm{V}$. The drain-current magnitude, $|I_{\mathrm{DS}}|$, is plotted as a function of gate voltage, $V_{\mathrm{GS}}$. Upward and downward triangles correspond to the FWD and BWD gate-voltage sweeps, respectively.}
\label{fig:RT_IV}
\end{figure*}

Fig.~\ref{fig:RT_IV} shows representative transfer characteristics measured at $T=77~\mathrm{K}$ and room temperature. The room-temperature data were acquired under vacuum using the TIA-based measurement set-up. By contrast, the $77~\mathrm{K}$ measurement was performed without vacuum using a two-SMU configuration, with one SMU applying $V_{\mathrm{GS}}$ and monitoring $I_{\mathrm{GS}}$, and the second applying $V_{\mathrm{DS}}$ and measuring $I_{\mathrm{DS}}$. For both temperatures, the FWD and BWD gate-voltage sweeps are shown using upper pointed and downward pointed triangles, respectively.

The room-temperature transfer characteristic exhibits substantially greater hysteresis than the $77~\mathrm{K}$ measurement, consistent with increased thermally activated trapping and detrapping at elevated temperature~\cite{lan2020origin,park2016thermally}. As no measurements were performed between $77~\mathrm{K}$ and room temperature, the temperature at which this increase in hysteresis develops cannot be determined. Nevertheless, the comparatively small hysteresis observed at $77~\mathrm{K}$, despite the measurement being conducted outside vacuum, suggests that the suppression is not solely due to the measurement environment. Instead, it is likely to be governed primarily by temperature-dependent charge-trapping kinetics at the WS$_2$/SiO$_2$ interface and associated defect states~\cite{withers2014electron,iqbal2015high,jana2023blocking}.

\begin{figure*}[h]
\centering
\includegraphics[width=0.6\textwidth]{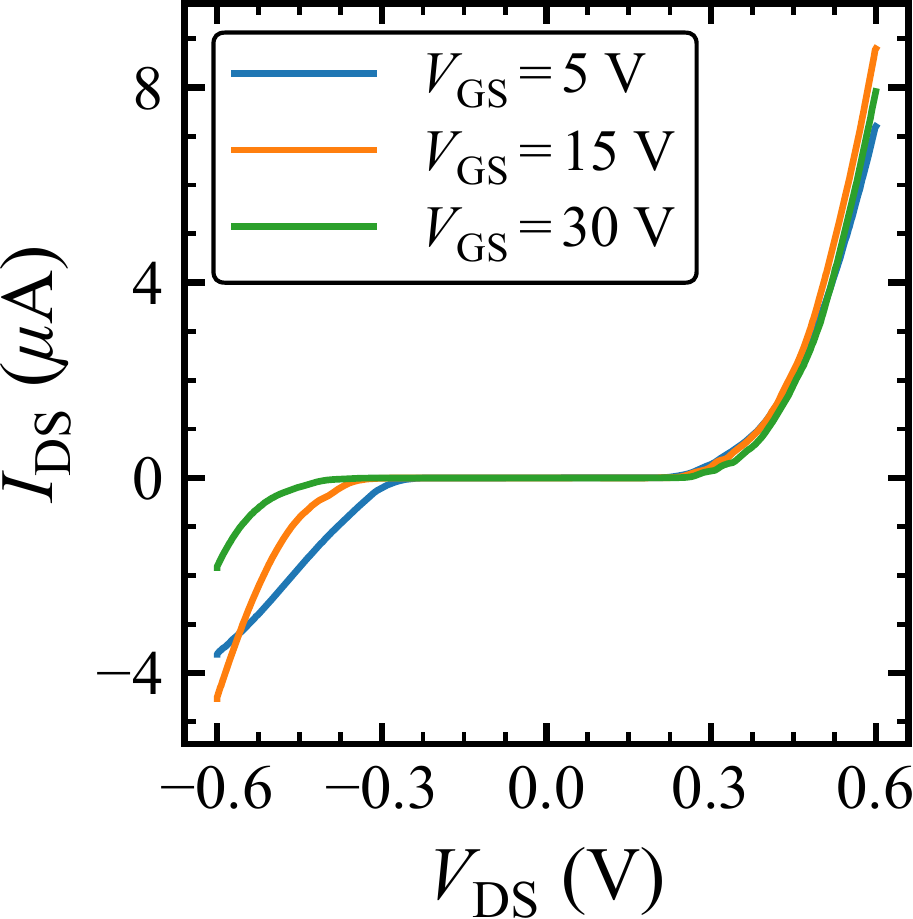}
\caption{Output characteristics of the back-gated WS$_2$ FET taken at $T = 20$ mK measured at selected gate voltages, $V_{\mathrm{GS}} = 5$, $15$, and $30~\mathrm{V}$. The drain current, $I_{\mathrm{DS}}$, is plotted as a function of drain-source voltage, $V_{\mathrm{DS}}$, over the full measured bias range. The data show an asymmetric current response for opposite drain-source bias polarities.}
\label{fig:output}
\end{figure*}

Fig~\ref{fig:output} shows the output characteristics of the DUT at three representative gate-voltage values. These data highlight the asymmetric current response under opposite drain-source bias polarities, consistent with asymmetric source/drain contact behaviour. The observed bias asymmetry is consistent with bias-dependent modification of the Schottky barrier at the injecting contact, including field-induced barrier lowering and barrier thinning. While contact- and channel-related transport mechanisms cannot be fully disentangled from the present measurements, the behaviour can be described more generally in terms of non-ideal and potentially asymmetric effective carrier injection \cite{Liu2014MoS2SchottkyBarrierTransistors,DiBartolomeo2018SchottkyMoS2}. Such behaviour is plausible for the present non-optimised contact geometry, since achieving low-temperature ohmic injection in TMDC devices typically requires deliberate contact engineering, for example through hBN-encapsulated heterostructures with etched edge contacts. Without such an architecture, the metal--WS$_2$ interfaces are expected to retain Schottky-like character at cryogenic temperatures \cite{Xu2016UniversalLowTOhmicContacts}. Bias-asymmetric transport has similarly been described in TMDC FETs using back-to-back Schottky-barrier models, while negative-transconductance-like behaviour in multilayer TMDCs has been associated with barrier-controlled vertical or interlayer transport. The measured response may therefore reflect contributions from more than one of these mechanisms \cite{Liu2014MoS2SchottkyBarrierTransistors,DiBartolomeo2018SchottkyMoS2,Liu2017VerticalNTC,Chen2021LowTNTC}.

\begin{figure*}[h]
\centering
\includegraphics[width=\textwidth]{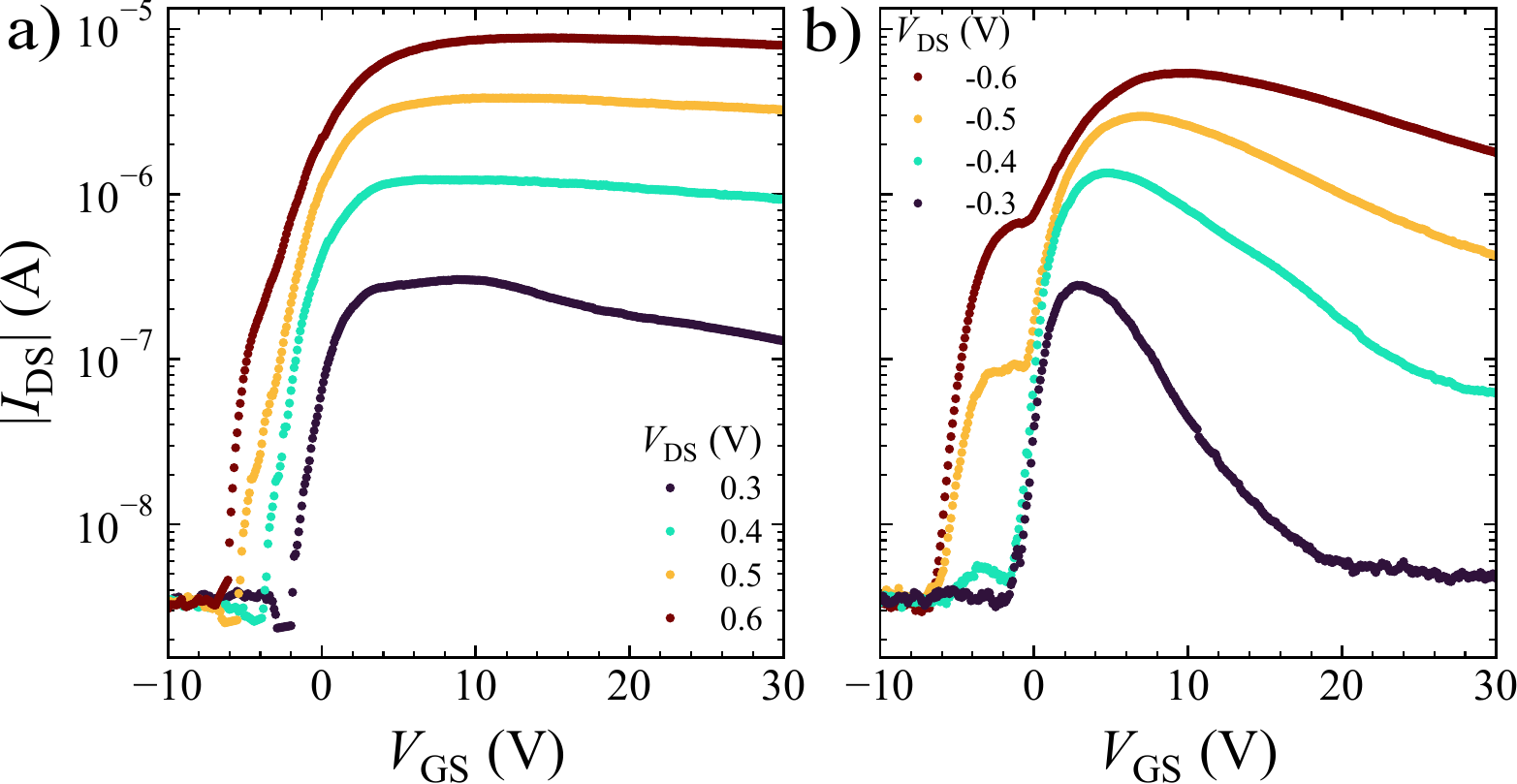}
\caption{Transfer characteristics of the DUT measured at \(T = 20~\mathrm{mK}\) for multiple drain--source biases, shown for the FWD gate-voltage sweep only. Panels (a) and (b) show \(\left|I_{\mathrm{DS}}\right|\) as a function of \(V_{\mathrm{GS}}\) on a logarithmic current scale for positive and negative displayed \(V_{\mathrm{DS}}\), respectively. For negative displayed \(V_{\mathrm{DS}}\), panel (b), the transfer curves show a pronounced reduction in current at large positive \(V_{\mathrm{GS}}\), corresponding to a negative transconductance response. Due to the lower TIA gain used for this measurement, the off-state current approaches the current-resolution limit of the measurement setup rather than reflecting the intrinsic DUT response.}
\label{fig:20mK_Transfer}
\end{figure*}

Fig~\ref{fig:20mK_Transfer} shows transfer characteristics measured at $T = 20~\mathrm{mK}$ for opposite source/drain bias polarities, plotted on both linear and logarithmic current scales. A more pronounced negative transconductance response is observed for the negative $V_{\mathrm{DS}}$ polarity. The measurements were performed using a lower TIA gain to capture the on-state behaviour of the DUT. As a result, the off-state regime is limited by the finite current resolution of the measurement setup, so the lowest measured currents should be interpreted as an upper bound on the intrinsic device current.

\clearpage

\section{Fitting Multi-branch Characteristics}
\label{app:multi-fit}

\begin{figure*}[h]
\centering
\includegraphics[width=0.7\textwidth]{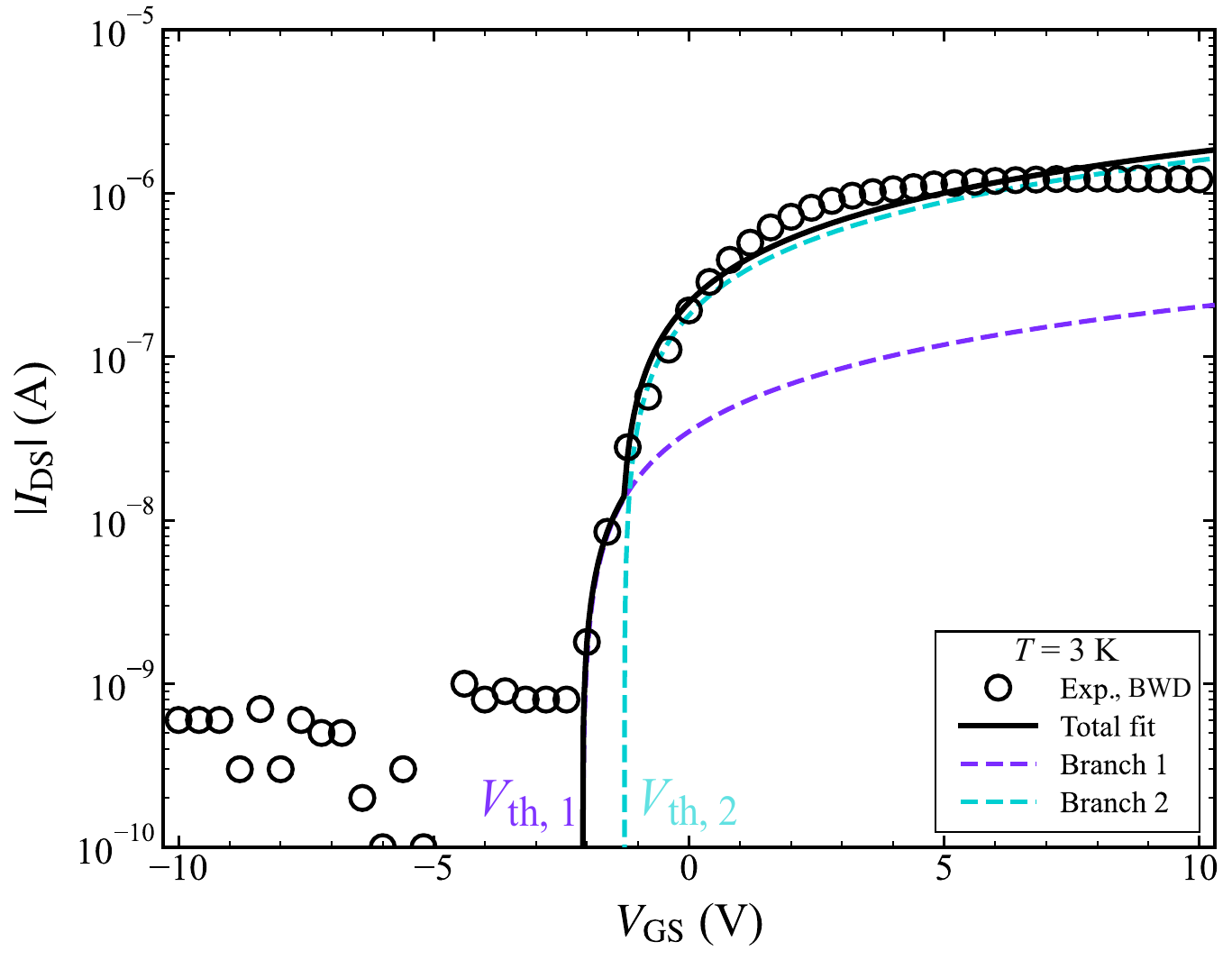}
\caption{Example of the multi-branch conduction fit to the BWD-sweep transfer characteristic measured at $V_{DS}=0.4~\mathrm{V}$ and $T=3~\mathrm{K}$. Open circles show the measured $I_{DS}$--$V_{GS}$ data, while the solid line shows the total fit obtained from the sum of the individual conduction branches. Dashed lines indicate the separate transistor branches, each with its own threshold voltage $V_{\mathrm{th}}$.}
\label{fig:3K_BWD}
\end{figure*}

\begin{figure*}[h]
\centering
\includegraphics[width=0.7\textwidth]{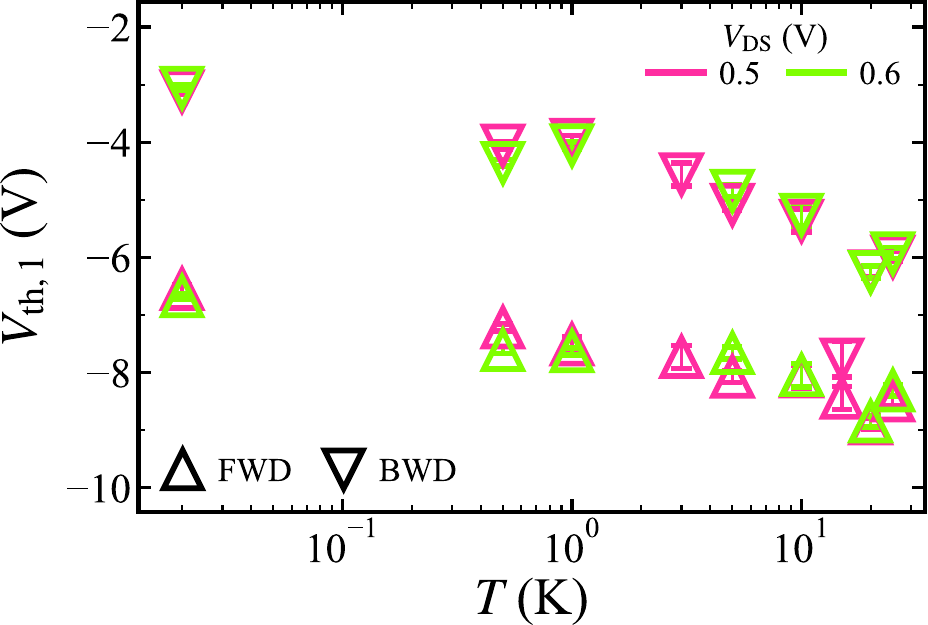}
\caption{Temperature dependence of the first extracted branch threshold voltage, $V_{\mathrm{th},1}$, for the FWD and BWD gate-voltage sweeps at $V_{\mathrm{DS}}=0.5$ and $0.6~\mathrm{V}$. FWD and BWD data are represented by upward- and downward-pointing triangles, respectively. Error bars indicate the $\pm 1\sigma$ uncertainty obtained from the fitting model.}
\label{fig:extra_vth}
\end{figure*}

Fig~\ref{fig:3K_BWD} provides a complementary example to the 3 K FWD-sweep fit shown in Fig~\ref{fig:lambertw_fit}(a), demonstrating the multi-branch fit of the 3 K BWD-sweep. Fig~\ref{fig:extra_vth} further supports the temperature-dependent trend in $V_{\mathrm{th},1}$ shown in Fig~\ref{fig:lambertw_fit}(b), highlighting the reproducibility of the extracted behaviour for both sweep directions and at $V_{DS}=0.5$ and $0.6~\mathrm{V}$.

The transfer characteristics were fitted using the one-, two- and three-branch versions of the model described in the main text, see Eq.~\ref{eq:lambertw_multibranch}. For each dataset, the candidate models were fitted independently and compared using the model-selection criteria described below. After fitting, the extracted branches were ordered by increasing threshold voltage, such that branch 1 corresponded to the earliest effective turn-on, $V_{\mathrm{th},1}$, branch 2 to the next turn-on, $V_{\mathrm{th},2}$, and branch 3 to the latest turn-on, $V_{\mathrm{th},3}$. This ordering was applied only after fitting and was used to give consistent branch labels across the temperature-dependent data.

The fit quality was assessed using both logarithmic and linear residuals. For the logarithmic fit metric, the measured and fitted currents were first transformed into log-current space as
\begin{equation}
y_i = \log_{10}\left(I_{i,\mathrm{data}}\right)
\end{equation}
and
\begin{equation}
\hat{y}_i = \log_{10}\left(I_{i,\mathrm{fit}}\right),
\end{equation}
where $y_i$ is the log-transformed measured current at the $i$-th fitted data point and $\hat{y}_i$ is the corresponding log-transformed fitted current. The logarithmic coefficient of determination was then calculated as
\begin{equation}
R^{2}_{\log} =
1 -
\frac{
\sum_{i=1}^{N} \left(y_i-\hat{y}_i\right)^{2}
}{
\sum_{i=1}^{N} \left(y_i-\bar{y}\right)^{2}
},
\end{equation}
where
\begin{equation}
\bar{y} = \frac{1}{N}\sum_{i=1}^{N} y_i
\end{equation}
is the mean of the log-transformed measured current values and $N$ is the number of fitted data points.

\begin{figure*}[h]
\centering
\includegraphics[width=\textwidth]{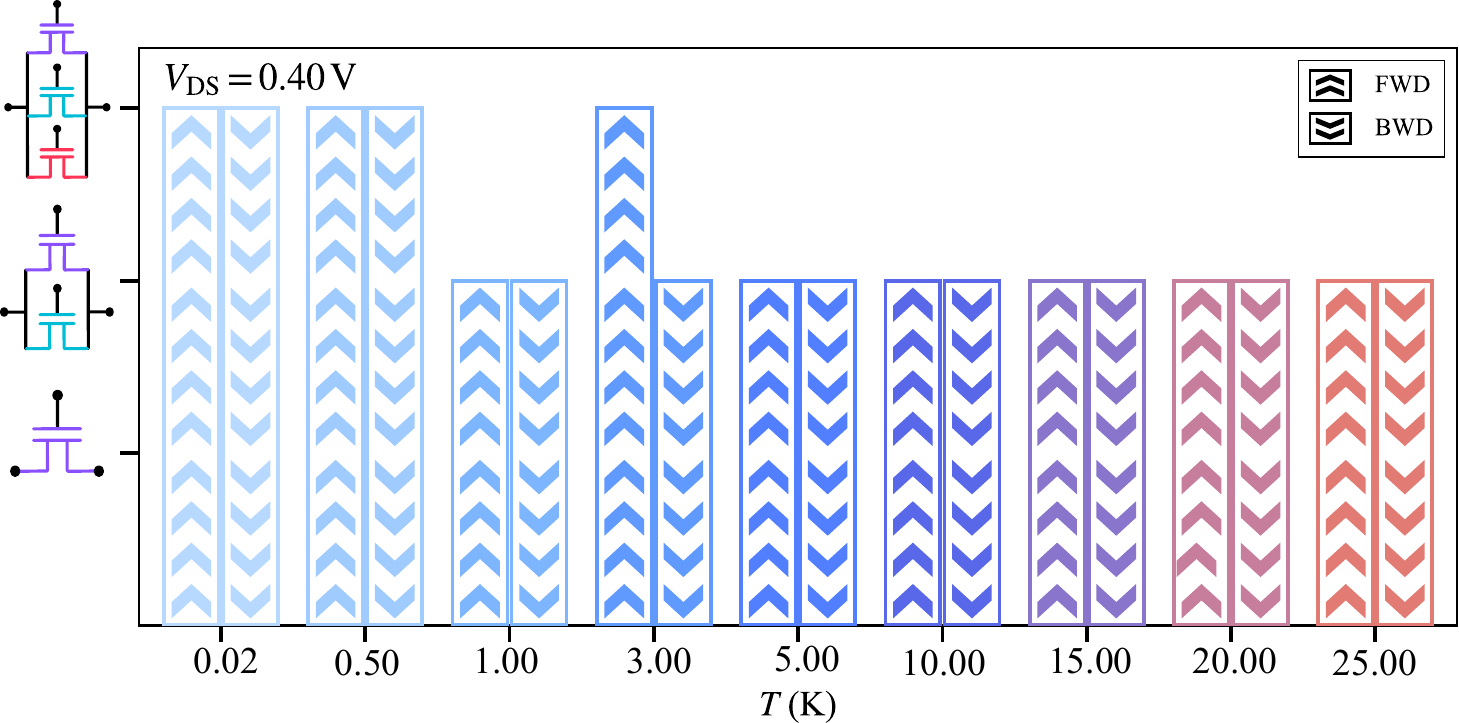}
\caption{Branch-selection map for the multi-branch Lambert-W fits at $V_{\textup{DS}} = 0.4$ V. The columns show the measurement temperature, while the rows indicate the schematic transport-branch configuration selected by the fitting procedure. Triangle markers indicate the selected configuration for each temperature and sweep direction, with upward and downward triangles corresponding to the FWD and BWD gate-voltage sweeps, respectively. The schematic circuits are used only as a visual representation of the number of effective transport branches.}
\label{fig:branch_selection_map}
\end{figure*}

The corrected Akaike information criterion, AICc, was used to compare the independently fitted two- and three-branch models. This allowed the three-branch model to be selected only when the improvement in fit quality was sufficient to justify the additional fitted parameters introduced by the extra branch~\cite{Sato2008OptimalPolynomialAICc}. For least-squares fitting, AICc was calculated as

\begin{equation}
\mathrm{AICc}_{n_{\textup{branch}}}
=
N\ln\left(\frac{\mathrm{RSS}}{N}\right)
+
2p
+
\frac{2p(p+1)}{N-p-1},
\end{equation}

where $N$ is the number of fitted data points, $p$ is the number of fitted parameters, and $\mathrm{RSS}=\sum_{i=1}^{N}\left(y_i-\hat{y}_i\right)^2$ is the residual sum of squares calculated in log-current space.
Model selection was performed successively, first by comparing the one- and two-branch models and then, where the two-branch model was preferred, by comparing the two- and three-branch models. In each comparison, the model containing the additional branch was retained only when the improvement in fit quality was sufficient to justify the additional fitted parameters and the extracted threshold voltages satisfied the minimum-separation criterion defined below. The three-branch model was retained only when it gave a sufficient improvement over the two-branch model, defined by
\begin{equation}
\mathrm{AICc}_{2}-\mathrm{AICc}_{3}>4,
\end{equation}
or
\begin{equation}
R^{2}_{\log,3}-R^{2}_{\log,2}>0.010.
\end{equation}
In addition, the extracted branch threshold voltages were required to be separated by at least four steps of the resolution of the gate voltage sweep, leading to
\begin{equation}
\min_{m\neq \ell}\left|V_{\mathrm{th},m}-V_{\mathrm{th},\ell}\right|
\geq
4\Delta V_{\mathrm{GS,step}}.
\end{equation}
If these criteria were not satisfied, the two-branch model was selected. A detailed map of the branch selection across the temperature range is shown in Fig~\ref{fig:branch_selection_map}.

\begin{figure*}[h]
\centering
\includegraphics[width=\textwidth]{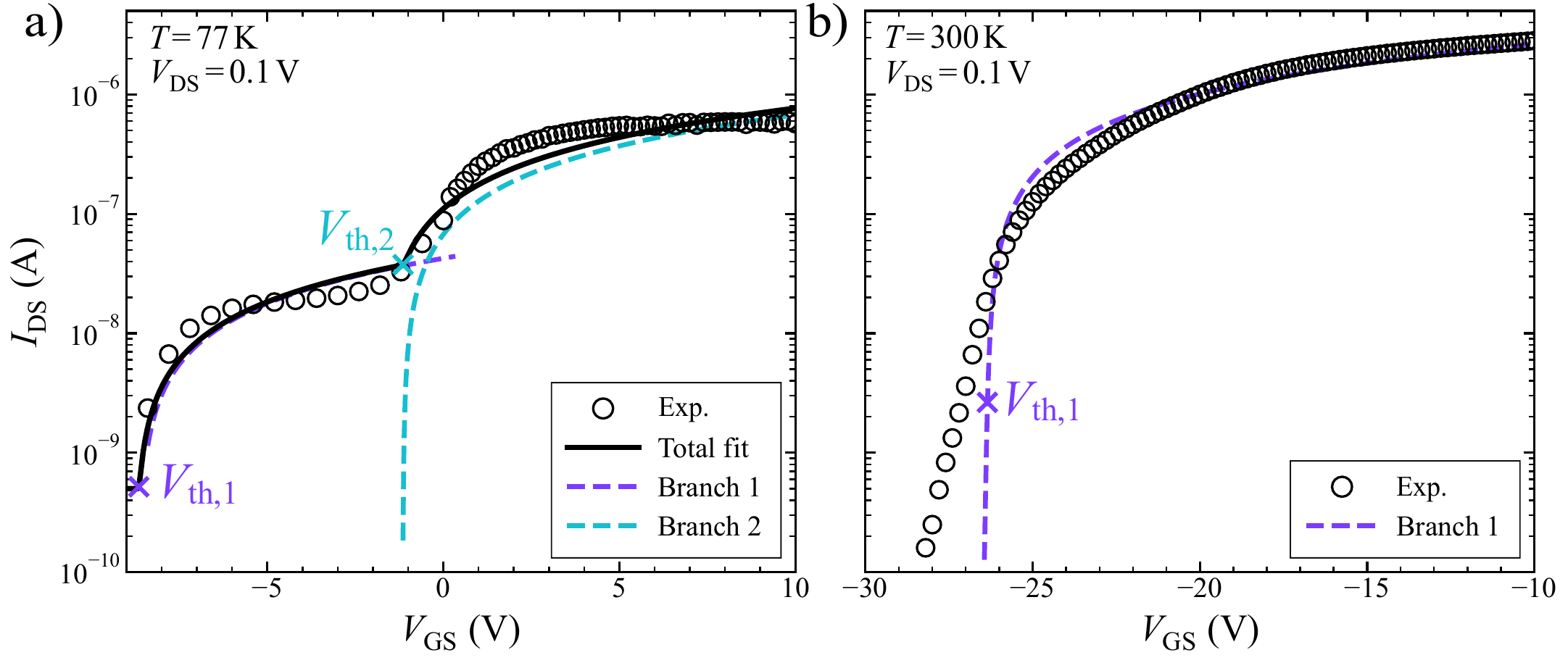}
\caption{Examples of the multi-branch conduction fits to the FWD-sweep transfer characteristics measured at $V_{DS}=0.1~\mathrm{V}$ for (a) $T=77~\mathrm{K}$ and (b) $T=300~\mathrm{K}$. These examples correspond to $n_{\textup{branch}}=2$ and $n_{\textup{branch}}=1$, respectively. Open circles show the measured $I_{DS}$--$V_{GS}$ data, while solid lines show the total fit obtained from the sum of the individual conduction branches. Dashed lines indicate the separate transistor branches, each with its own threshold voltage $V_{\mathrm{th}}$, illustrating how multiple turn-on channels contribute to the overall transfer curve.}
\label{fig:MultiTransistor_ExtendedFits}
\end{figure*}

Fig~\ref{fig:MultiTransistor_ExtendedFits} presents examples of a two-branch fit at 77 K and a one-branch fit at room temperature. At room temperature, a single branch is sufficient because no clear shoulder-like structure is observed in the transfer characteristic. Moreover, for the room-temperature data, a deviation from the ideal logarithmic turn-on is observed at lower current. This is attributed to finite series resistance in the DC measurement wiring of the dilution refrigerator loom. As a result, the measured transfer characteristic is not expected to follow the ideal field-effect response over the full current range. This contribution could be reduced using a Kelvin four-terminal measurement configuration, which would separate the voltage dropped across the device from voltage drops in the measurement wiring.

\clearpage

\section{Computational Methods}
\label{app:comp_methods}

The electronic-structure calculations were performed within density functional theory (DFT) using linear-combination-of-atomic-orbitals (LCAO) basis sets, as implemented in QuantumATK \cite{smidstrup2020quantumatk}. Norm-conserving pseudopotentials from the PseudoDojo library were used in combination with medium-quality basis sets \cite{van2018pseudodojo}. Brillouin-zone integration was performed using a Monkhorst–Pack k-point sampling scheme \cite{monkhorst1976special}, with a sampling density of approximately 10 k-points per reciprocal A$^{\circ}$. A real-space grid energy cutoff of 115 Hartree was applied. Exchange and correlation were described using the Perdew–Burke–Ernzerhof (PBE) functional within the generalised-gradient approximation (GGA), while van der Waals interactions were included using the Grimme DFT-D3 dispersion correction \cite{grimme2010consistent}. To suppress spurious interactions between neighbouring periodic images, a 20 $\AA$ vacuum region was included along the out-of-plane (z) direction. Calculations were performed at zero field and under a perpendicular electric field of 0.5 V/nm, applied normal to the WS$_2$ layers. Each structure was independently relaxed under its respective electric-field condition using the limited-memory Broyden–Fletcher–Goldfarb–Shanno algorithm until the maximum force on any atom was below 0.02 eV/A$^{\circ}$. Spin–orbit coupling was not included in the calculations.

\subsection{Field-induced layer redistribution}
\label{sec:field_induced_layer_redistribution}

To assess whether perpendicular gating can produce electronically distinct conduction contributions within a multilayer WS$_2$ channel, layer-projected band structures were calculated for slabs containing three to six layers. At zero electric field, the valence-band maximum (VBM) and conduction-band minimum (CBM) are distributed approximately symmetrically across the slab. Their layer-polarization values, $P_z$, remain close to zero for every thickness considered, while the effective number of contributing layers, $N_{\mathrm{eff}}$, increases with the number of layers. This indicates that the zero-field band-edge states are delocalized over a substantial fraction of the multilayer stack rather than being confined to either outer surface.

\begin{figure*}[h]
\centering
\includegraphics[width=\textwidth]{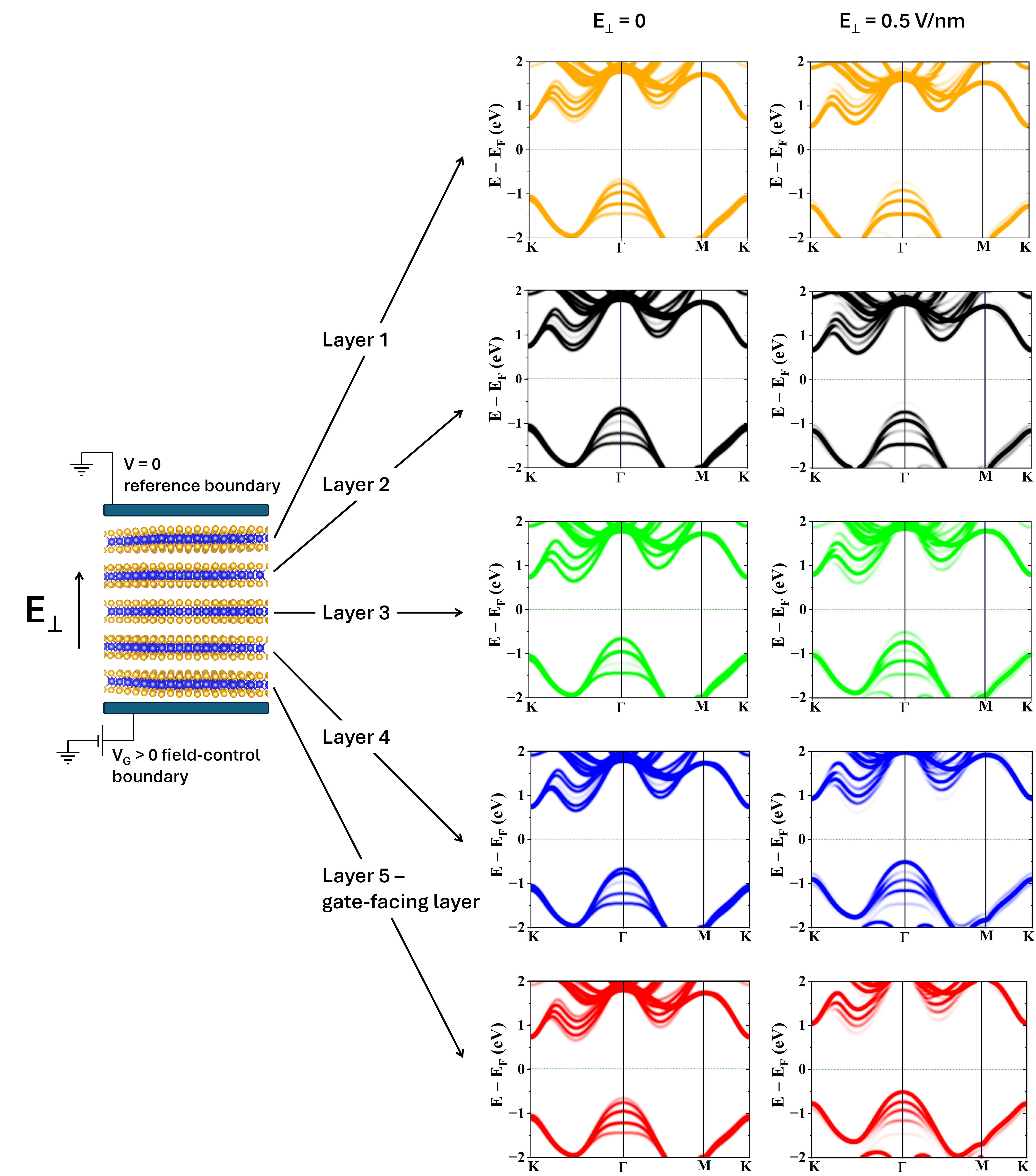}
\caption{Electric-field-induced redistribution of the band-edge states in five-layer WS$_2$. The left and right columns show layer-projected band structures calculated at E$_{\perp}$ = 0 and E$_{\perp}$ = 0.5 V/nm, respectively. The five rows correspond to Layers 1-5, as indicated by the atomic-structure schematic, with Layer 5 representing the gate-facing layer. The upper grounded and lower biased planes schematically represent the electrostatic boundary conditions used to indicate the direction of the applied perpendicular field. The colour intensity represents the relative projection weight on each layer. At zero field, the band-edge states are distributed across several layers, whereas the applied field produces pronounced layer asymmetry and strong localisation of the conduction-band minimum toward one outer layer. The energy zero denotes the Fermi level.}
\label{fig:dft_calc}
\end{figure*}

Application of a perpendicular electric field of $0.5~\mathrm{Vnm^{-1}}$ produces a pronounced redistribution of the band-edge states, as shown in Fig.~\ref{fig:dft_calc}. For the four-, five-, and six-layer slabs, the CBM becomes almost completely localized on one outer layer, giving $\lvert P_z\rvert \simeq 1$ and $N_{\mathrm{eff}} \simeq 1$. The VBM shifts towards the opposite side of the slab but remains distributed over approximately two to three layers. The magnitude of the VBM--CBM polarization difference increases with thickness, reaching $\lvert\Delta P_z\rvert = 1.51$, $1.64$, and $1.73$ for the four-, five-, and six-layer structures, respectively.

The three-layer calculation also develops strong band-edge asymmetry under the applied field, although the sign of $P_z$ differs from those of the thicker structures because the CBM localizes on the oppositely indexed surface. The sign depends on the chosen field direction and layer-numbering convention; therefore, the physically relevant measure of the band-edge separation is $\lvert\Delta P_z\rvert$. Across all thicknesses, the calculations demonstrate that a perpendicular electric field can make nominally equivalent WS$_2$ layers electronically inequivalent and can spatially separate the VBM and CBM. This result does not establish a one-to-one correspondence between the experimentally fitted transport branches and individual atomic layers. It does, however, provide an atomistic mechanism through which different regions of a multilayer channel may acquire different effective turn-on conditions under back-gate bias, supporting field-induced layer-dependent electrostatics as one possible contribution to the observed multi-stage turn-on behaviour. As the calculations employ ideal three- to six-layer slabs, whereas the experimental channel is approximately 90 nm thick and is expected to exhibit substantially stronger electrostatic screening, the calculations are intended to establish a qualitative microscopic mechanism rather than to reproduce quantitatively the electrostatics of the measured device. 

For each band-edge eigenstate, the orbital-projection weight on layer $l$, denoted $w_l$, was extracted from the layer-projected band structure and normalized according to
\begin{equation}
P_l =
\frac{w_l}{\displaystyle\sum_j w_j},
\qquad
\sum_l P_l = 1.
\label{eq:layer_projection_probability}
\end{equation}

The layer polarization was calculated as
\begin{equation}
P_z =
\sum_{l=1}^{N} \xi_l P_l,
\label{eq:layer_polarization}
\end{equation}
where $\xi_l$ varies linearly from $-1$ for the lowest layer to $+1$ for the highest layer. For the layer-numbering convention used in Fig.~\ref{fig:dft_calc}, Layer~5 is the lowest, gate-facing layer and Layer~1 is the highest layer. Consequently, $P_z=0$ represents a state that is centred within, or symmetrically distributed across, the slab, whereas $P_z=-1$ and $P_z=+1$ represent complete localization on the two opposite outer layers.

The band-edge spatial separation was quantified as
\begin{equation}
\Delta P_z =
P_z^{\mathrm{CBM}} -
P_z^{\mathrm{VBM}}.
\label{eq:band_edge_polarization_separation}
\end{equation}

The degree of delocalization across the stack was quantified using the effective number of contributing layers,
\begin{equation}
N_{\mathrm{eff}} =
\frac{1}{\displaystyle\sum_l P_l^2}.
\label{eq:effective_layer_number}
\end{equation}
Here, $N_{\mathrm{eff}}=1$ corresponds to complete localization on one layer, whereas $N_{\mathrm{eff}}=N$ corresponds to an equal distribution over all $N$ layers.

\begin{table*}[htbp]
\centering
\caption{Layer-resolved electronic properties of three- to six-layer WS$_2$ slabs calculated at zero perpendicular electric field and at $E_{\perp}=0.5~\mathrm{Vnm^{-1}}$. The quantities $P_z^{\mathrm{VBM}}$ and $P_z^{\mathrm{CBM}}$ describe the normalized positions of the valence- and conduction-band-edge states across the multilayer stack, with $P_z=-1$ and $P_z=+1$ corresponding to localization on opposite outer layers. The quantity $N_{\mathrm{eff}}$ gives the effective number of layers contributing to each state. The band-edge separation is defined as $\Delta P_z=P_z^{\mathrm{CBM}}-P_z^{\mathrm{VBM}}$, and its absolute magnitude, $\lvert\Delta P_z\rvert$, is reported.}
\label{tab:dft}

\normalsize
\renewcommand{\arraystretch}{1.20}

\begin{tabular*}{\textwidth}
{@{\extracolsep{\fill}} c c cc cc c @{}}
\toprule

Number of layers
& $E_{\perp}$
& $P_z^{\mathrm{VBM}}$
& $N_{\mathrm{eff}}^{\mathrm{VBM}}$
& $P_z^{\mathrm{CBM}}$
& $N_{\mathrm{eff}}^{\mathrm{CBM}}$
& $\lvert\Delta P_z\rvert$
\\

&
($\mathrm{V\,nm^{-1}}$)
&
&
&
&
&
\\

\midrule

3
& 0.0
& $-0.005$
& 2.35
& 0.002
& 2.71
& 0.007
\\

3
& 0.5
& 0.229
& 2.37
& $-1.000$
& 1.00
& 1.229
\\

4
& 0.0
& $-0.006$
& 3.23
& 0.007
& 3.52
& 0.012
\\

4
& 0.5
& $-0.506$
& 2.45
& 1.000
& 1.00
& 1.506
\\

5
& 0.0
& 0.008
& 3.97
& $-0.007$
& 4.09
& 0.015
\\

5
& 0.5
& $-0.636$
& 2.49
& 1.000
& 1.00
& 1.636
\\

6
& 0.0
& $-0.021$
& 5.16
& 0.015
& 5.14
& 0.037
\\

6
& 0.5
& $-0.726$
& 2.35
& 1.000
& 1.00
& 1.726
\\

\bottomrule

\end{tabular*}

\end{table*}

\begin{figure*}[h]
\centering
\includegraphics[width=\textwidth]{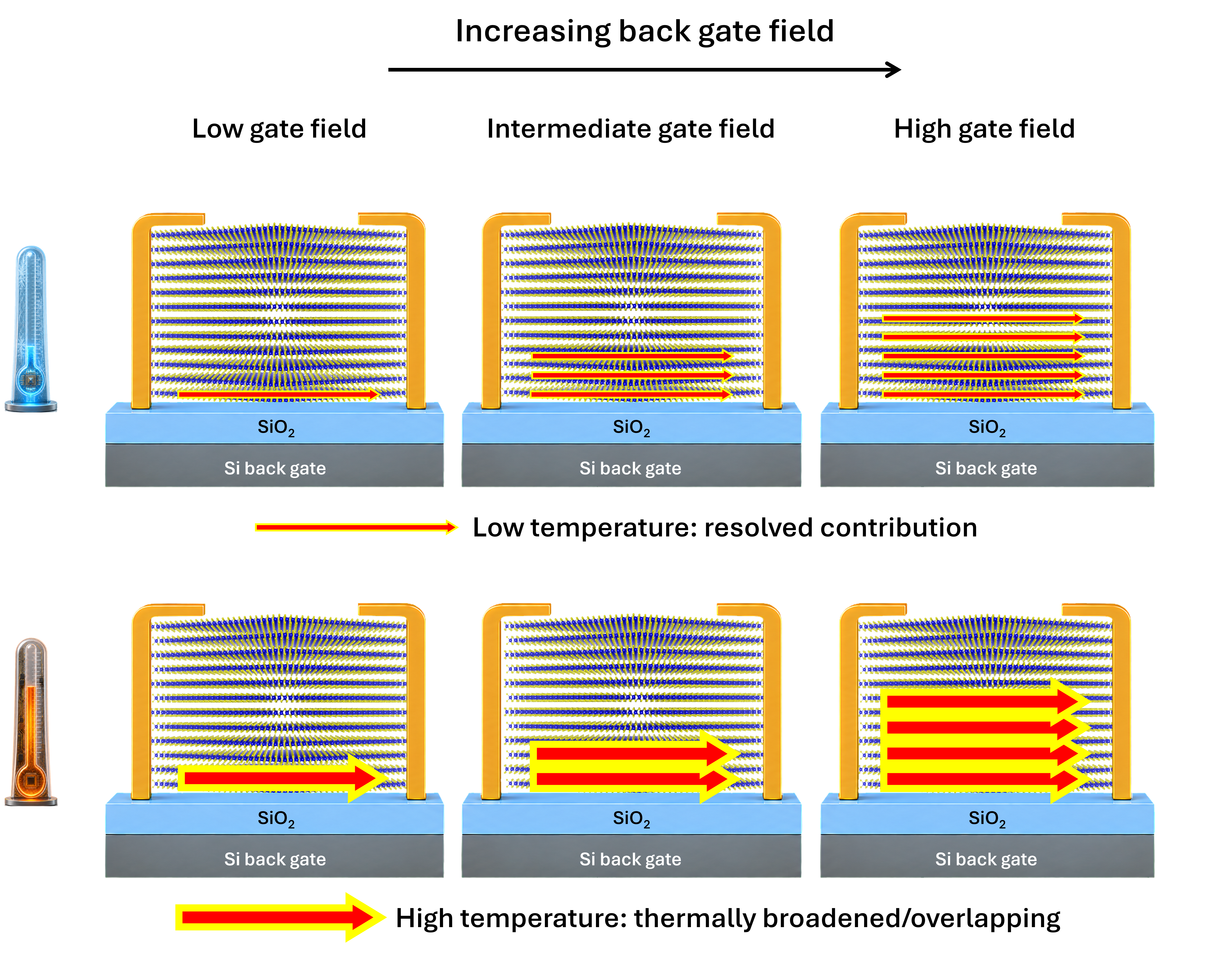}
\caption{Schematic interpretation of gate- and temperature-dependent multi-branch transport in multilayer WS$_2$. From left to right, increasing back-gate field progressively activates additional effective lateral conduction contributions. At low temperature (top row), these contributions remain comparatively narrow and may be resolved as separate turn-on features. At higher temperature (bottom row), thermal activation and broadening cause neighbouring contributions to overlap, so fewer effective branches are distinguishable in the measured transfer characteristics or required by the model. Arrow direction indicates lateral source-to-drain transport. The arrow positions are illustrative and do not imply a one-to-one correspondence between fitted branches and individual WS$_2$ layers.}
\label{fig:schematic_multilayer}
\end{figure*}

This interpretation is summarized schematically in Fig.~\ref{fig:schematic_multilayer}. At low temperature, the effective transport contributions remain sufficiently narrow in gate voltage to produce distinct shoulders in the transfer characteristics. Increasing the back-gate field progressively activates additional current-carrying regions within the multilayer channel. At higher temperature, thermal activation and broadening cause neighbouring contributions to overlap, so fewer effective branches can be distinguished experimentally or required by the fit. The illustrated pathways are phenomenological effective contributions and should not be assigned directly to individual WS$_2$ layers.

\clearpage

\twocolumngrid
\bibliography{ref_main}

\end{document}